\documentclass[11pt,a4paper]{article}

\usepackage{amsmath}
\usepackage[T1]{fontenc}
\usepackage[normalem]{ulem}

\usepackage{jheppub}
\usepackage{psfrag}
\usepackage{slashed}
\usepackage{cancel}
\usepackage{lscape}

\usepackage{caption}
\usepackage{array}
\usepackage{graphicx}
\usepackage{subcaption}
\usepackage{multirow}
\usepackage{tabularx}
\usepackage{makecell}
\usepackage{xspace}
\usepackage[table]{xcolor}

\usepackage{color, soul}

\usepackage[colorinlistoftodos]{todonotes}
\setuptodonotes{size=\scriptsize}
\usepackage[utf8]{inputenc}

\renewcommand{\arraystretch}{1.3}

\usepackage[utf8]{inputenc}
\usepackage{graphicx}
\usepackage{slashed}
\usepackage{color}
\usepackage{amsmath}
\usepackage{amssymb}

\definecolor{mypink}{RGB}{219, 48, 122}
\definecolor{mygreen}{rgb}{0,0.7,0}
\definecolor{byzantine}{rgb}{0.74, 0.2, 0.64}

\def\hpl11{{\mathrm{HPL}}_{1,1}}

\newcolumntype{C}[1]{>{\hsize=#1\hsize\centering\arraybackslash}X}%

\newcolumntype{Z}{r<{\hspace{3mm}}}

\def\bbbar{\ensuremath{b\kern1pt\bar{b}}\xspace}
\def\wbb{\ensuremath{\operatorname{W}\kern-1pt \bbbar}\xspace}
\def\wbbj{\ensuremath{\operatorname{W}\kern-1pt \bbbar j}\xspace}
\def\wpbb{\ensuremath{\operatorname{W}^+\kern-1pt \bbbar}}
\def\wmbb{\ensuremath{\operatorname{W}^-\kern-1pt \bbbar}}
\def\rT{\ensuremath{\text{T}}}
\def\kt{\ensuremath{k_\text{T}}\xspace}
\def\pt{\ensuremath{p_\text{T}}\xspace}
\def\Pp{\ensuremath{\operatorname{p}\kern-1pt}}
\def\PW{\ensuremath{\operatorname{W}\kern-1pt}}
\def\PWp{\ensuremath{\mathrm{W^+}}\xspace}
\def\PWm{\ensuremath{\mathrm{W^-}}\xspace}
\def\PZ{\ensuremath{\operatorname{Z}\kern-1pt}}
\def\PH{\ensuremath{\operatorname{H}\kern-1pt}}
\def\RT{\ensuremath{\text{T}}}

\def\cK{K}

\newcommand\GeV{\ensuremath{\mathrm{GeV}}\xspace}

\def\lo{\mathrm{LO}}
\def\nlo{\mathrm{NLO}}
\def\nloplus{\mathrm{NLO+}}
\def\nnlo{\mathrm{NNLO}}
\def\gev{\mathrm{GeV}}
\def\inc{\mathrm{inc}}
\def\exc{\mathrm{exc}}

\title{Flavour anti-$k_\text{T}$ algorithm applied to $\wbb$ production at the LHC}

\author[a]{Heribertus Bayu Hartanto,}
\author[a]{Rene Poncelet,}
\author[a]{Andrei Popescu,}
\author[b]{Simone Zoia}

\affiliation[a]{
Cavendish Laboratory, University of Cambridge, Cambridge CB3 0HE, United Kingdom
}
\affiliation[b]{
Dipartimento di Fisica and Arnold-Regge Center, Università di Torino, and INFN, Sezione di
Torino, Via P. Giuria 1, I-10125 Torino, Italy
}

\emailAdd{
hbhartanto@hep.phy.cam.ac.uk,
poncelet@hep.phy.cam.ac.uk,
popescu@hep.phy.cam.ac.uk,
simone.zoia@unito.it
}

\abstract{
We apply the recently proposed flavoured anti-$k_{\text{T}}$ jet algorithm to $\wbb$ production at
the Large Hadron Collider at $\sqrt{s}=8$ TeV.
We present results for the total cross section and differential distributions at the next-to-next-to-leading order (NNLO) in QCD.
We discuss the effects of the remaining parametric freedom in the flavoured anti-\kt prescription, and compare it against the standard flavour-\kt algorithm.
We compare the total cross section results against the CMS data, finding good agreement.
The NNLO QCD corrections are significant, and their inclusion substantially improves the agreement with the data.
}

\preprint{CAVENDISH-HEP-22/07}

\begin{document}
\maketitle
\flushbottom
\section{Introduction}
\label{sec:introduction}

The Large Hadron Collider (LHC) has recently started its Run 3 phase, following the successful Run 1 and 2 campaigns.
The LHC dataset collected from the previous runs, combined with the current Run 3 and the upcoming
high-luminosity LHC (HL-LHC) phase, will provide opportunities to stress test the Standard Model of Particle Physics (SM) and constrain new physics scenarios.
To keep pace with the increasingly accurate experimental measurements, theory predictions must be provided with the highest possible precision.
One of the current precision frontiers at the LHC is the next-to-next-to-leading order (NNLO) in QCD for $2\to 3$ scattering processes.
The last few years have seen a major breakthrough in the massless $2\to 3$ computations, where NNLO
QCD predictions for $pp \to \gamma\gamma\gamma$~\cite{Kallweit:2020gcp,Chawdhry:2019bji}, $pp \to\gamma\gamma+$jet~\cite{Chawdhry:2021hkp,Badger:2021ohm} and, most notably, $pp\to$ 3-jet production~\cite{Czakon:2021mjy,Chen:2022ktf} have become available.
The first NNLO QCD calculation for $2\to 3 $ process involving one external mass has recently been
completed for the production of W-boson in association with a $b\bar{b}$ pair, where the W-boson decays leptonically and the bottom quark is treated as a massless particle~\cite{Hartanto:2022qhh}.
We call this process `$\wbb$ production' hereafter.

$\wbb$ production has been studied quite extensively, particularly at next-to-leading order (NLO) in QCD~\cite{Ellis:1998fv,FebresCordero:2006nvf,FebresCordero:2009xzo,Badger:2010mg,Oleari:2011ey,Frederix:2011qg,Luisoni:2015mpa,Anger:2017glm}.
The studies carried out at NLO QCD demonstrated that, for inclusive production (W in association with at least two $b$-jets in the final state), the higher-order corrections are significant and the theoretical uncertainty at NLO is larger than that of the LO prediction.
This pathological behaviour is attributed to the contribution of $qg$-initiated subprocesses, which start contributing at NLO.
The inclusion of the NNLO QCD corrections improves the perturbative convergence: the corrections become moderate, compared to NLO QCD, and the theoretical uncertainties appear to be under control.
Experimental measurements for $\wbb$ production have been undertaken at both the Tevatron~\cite{D0:2004prj} and the LHC~\cite{CMS:2013xis,CMS:2016eha}.

This interest in $\wbb$ production stems from multiple reasons.
First of all, the $\PW+2b$-jets signature is also produced by other SM scattering processes, such as
the $\PW+$Higgs-strahlung process, where the Higgs boson decays to a bottom-quark pair (i.e.\ $pp\to
\PW(\PH\to b\bar{b})$), and single top production (i.e.\ $pp \to (t\to \PW b)\bar{b}$).
The analysis of these important processes therefore requires good control of the irreducible background, represented by $\wbb$ production.
We note that the theoretical predictions for both the $\PW\PH\to b\bar{b})$ and single top productions are known through 
NNLO in QCD~\cite{Ferrera:2013yga,Campbell:2016jau,Ferrera:2017zex,Caola:2017xuq,Gauld:2019yng,Behring:2020uzq,Astill:2016hpa,Zanoli:2021iyp}.
Processes involving new physics may lead to the very same experimental signature as well.
$\wbb$ production is also a perfect testing ground to study the modelling of flavoured jets from both theoretical and experimental points of view.
An exclusive final state, where additional jets are vetoed, 
is often required in the experimental analyses to suppress the contributions from certain background processes.
The exclusive setup requires a more careful treatment of the theoretical uncertainties, as we already showed in Ref.~\cite{Hartanto:2022qhh}, where we compared the standard and the more conservative prescription of Ref.~\cite{Stewart:2011cf}.

The presence of $b$-quark-flavoured jets in the final state requires a flavour tagging procedure.
Experimentally, $b$-jets are tagged by identifying jets that contain a B-hadron by exploiting its long lifetime, looking for secondary vertices and tracks with a large impact parameter.
In theoretical simulations, the procedure depends on the type of calculation employed.
If a parton shower is involved, one can basically apply the experimental prescription, i.e.\ sequentially hadronise the showered event and put `$b$-jet' labels on the clustered jets which contain a B-hadron.
In a fixed-order calculation, the final state partons are assigned to jets by means of a clustering algorithm.

If the bottom quark is treated as a massless particle in a fixed-order calculation, care must be taken to ensure the infrared (IR) safety.
A problem might in fact arise from the assignment of flavoured particles which make up a soft pair to different clusters, thus spoiling the IR safety of the jet algorithm for the soft gluon emission.
This issue indeed sets in for the standard (flavour-blind) clustering algorithms, starting at NNLO.
We note that, if the bottom quark mass is instead taken into account, these IR divergences in the fixed-order calculation are regulated.

This problem can be remedied by employing a flavour-sensitive jet algorithm. The flavour-$k_\rT$ jet algorithm of Ref.~\cite{Banfi:2006hf} was proposed for this reason, and has been applied to several scattering processes at the LHC which involve final state bottom- or charm-quark jet(s)~\cite{Caola:2017xuq,Ferrera:2017zex,Gauld:2019yng,Gauld:2020ced,Gauld:2020deh,Czakon:2020coa,Hartanto:2022qhh}.
The major drawback in using the flavour-$k_\rT$ algorithm is that the experimental analyses at the LHC widely employ the standard anti-$k_\rT$ prescription~\cite{Cacciari:2008gp}.
A comparison of the flavour-$k_\rT$-based theoretical predictions to the experiments requires the experimental data to be unfolded from the standard anti-$k_\rT$ to the flavour-$k_\rT$ clustering, through the use of an NLO calculation matched to a parton shower (NLO+PS)~\cite{Gauld:2020deh}.
The unfolding procedure represents an additional source of uncertainty, which becomes sizeable when the clustering behaviours are significantly different.
In order to minimise the effect of unfolding, a flavour-sensitive anti-$k_\rT$ jet algorithm~\cite{Czakon:2022wam} has been recently introduced.
It matches more closely the standard anti-\kt prescription, and thus enables a more direct comparison with the experimental measurements at the LHC.
Recently, a new approach to dress anti-\kt jets with flavour has been proposed in Ref.~\cite{Gauld:2022lem}.

In this work we apply the newly proposed flavoured anti-$k_\text{T}$ jet algorithm to $\wbb$ production at NNLO QCD, compare with the results obtained using the flavour-$k_\rT$ algorithm, and compare the resulting cross sections to the measurements by the CMS collaboration~\cite{CMS:2016eha}.
This paper is structured as follows. In Section~\ref{sec:calculation} we briefly review the ingredients of the calculation.
Phenomenological results employing both variants of flavour sensitive jet algorithm are discussed in Section~\ref{sec:phenomenology}.
The genuine impact of NNLO QCD corrections is further studied in Section~\ref{sec:NLOplus} by comparing the NNLO calculation against an improved NLO prediction. We finally present our conclusions in Section~\ref{sec:conclusions}.


\section{Calculation and setup}
\label{sec:calculation}

We present the calculation of the $pp\to \PW(\to\ell\nu)b\bar{b}$ process at NNLO in QCD.
The computation has been performed within the \textsc{Stripper} framework, a \textsc{C++} implementation of the four-dimensional formulation of the sector-improved residue subtraction scheme~\cite{Czakon:2010td,Czakon:2014oma,Czakon:2019tmo}.
The tree-level matrix elements are supplied by the \textsc{AvH} library~\cite{Bury:2015dla}, while the one-loop matrix elements are provided by the \textsc{OpenLoops} package~\cite{Buccioni:2017yxi,Buccioni:2019sur}.
We computed the two-loop virtual amplitudes analytically in the leading colour approximation as described in Ref.~\cite{Hartanto:2022qhh}, following the strategy of Ref.~\cite{Badger:2021nhg}.
The resulting analytic expressions were implemented in \textsc{C++} for a fast numerical evaluation with the \textsc{PentagonFunctions++} library~\cite{Chicherin:2021dyp} as a backend for evaluating the special functions.

We work in the five-flavour scheme (5FS), where the bottom quark is treated as massless and is present in the initial state.
We present numerical results for the LHC center-of-mass energy $\sqrt{s}=8$ TeV.
We take the Cabbibo-Kobayashi-Maskawa (CKM) matrix to be diagonal and use the following input parameters:
\begin{align*}
&M_{\PW} = 80.351972 \; \gev, & \Gamma_{\PW} = 2.0842989 \; \gev, \nonumber \\
&M_{\PZ} = 91.153481 \; \gev, & \Gamma_{\PZ} = 2.4942665 \; \gev, \\
&G_F = 1.16638 \cdot 10^{-5} \;\gev^{-2}. & \nonumber
\end{align*}
The electromagnetic coupling $\alpha$ is determined within the $G_\mu$ scheme using the following relation:
\begin{equation}
\alpha = \frac{\sqrt{2}}{\pi} G_F M_{\PW}^2 \; \bigg(1-\frac{M_{\PW}^2}{M_{\PZ}^2}\bigg).
\end{equation}
We use the NNPDF3.1 PDF sets~\cite{NNPDF:2017mvq} via the LHAPDF interface~\cite{Buckley:2014ana}.
Specifically, we employ \verb=NNPDF31_lo_as_0118=, \verb=NNPDF31_nlo_as_0118=, \verb=NNPDF31_nnlo_as_0118=
for the leading order (LO), NLO and NNLO QCD calculations, respectively.

As we mentioned in Section~\ref{sec:introduction}, we employ both the flavour-sensitive $k_\rT$ and anti-$k_\rT$ jet algorithms, with the separation parameter $R=0.5$.
To define the fiducial region, we adopt the following selection cuts for
the jets and charged leptons~\cite{CMS:2016eha,Hartanto:2022qhh}:
\begin{align}
&p_{\RT,\ell} > 30 \;\gev,\qquad\qquad |\eta_{\ell}| < 2.1, \nonumber \\
&p_{\RT,j} > 25 \;\gev, \qquad\qquad |\eta_{j}| < 2.4 \label{eq:cuts},\\
&p_{\RT,b} > 25 \;\gev, \qquad\qquad |\eta_{b}| < 2.4 \nonumber.
\end{align}
In addition, we consider two different final state signatures:
\begin{itemize}
\item \textit{inclusive}: at least two $b$-jets are required in the final state;
\item \textit{exclusive}: exactly two $b$-jets and no other jets are required in the final state.
\end{itemize}
The jet veto parameter required to define the exclusive final state follows from Eq.~\eqref{eq:cuts}.

To obtain our predictions, we use a kinematic-dependent quantity for the renormalisation ($\mu_R)$ and factorisation ($\mu_F$) scales.
In particular, we utilise
\begin{equation}
H_\RT = E_\RT(\ell\nu) + p_\RT(b_1) + p_\RT(b_2)\,,
\label{eq:HTdef}
\end{equation}
where $E_\RT(\ell\nu) = \sqrt{M^2(\ell\nu)+p^2_\RT(\ell\nu)}$ is
the transverse energy of the lepton-neutrino system (i.e.\ the off-shell $\PW$ boson), $M(\ell\nu)$ and $p_\RT(\ell\nu)$ are the
corresponding invariant mass and transverse momentum, and $p_\RT(b_1)$ ($p_\RT(b_2)$) is the hardest (second hardest) bottom-quark-flavoured jet. We set $\mu_R = \mu_F = H_\RT$ for the central scale.

Before presenting our phenomenological results, let us briefly review the flavoured anti-\kt jet algorithm proposed in Ref.~\cite{Czakon:2022wam}.
The standard anti-\kt distance is multiplied by the following damping function $\mathcal S_{ij}$ if both pseudo-jets $i$ and $j$ have the same non-zero flavour of opposite sign:
\begin{align}
    \mathcal S_{ij} = 1 - \theta(1 - x) \, \cos \left(\frac\pi2 x\right) \le 1 \,,
\end{align}
where $\theta$ is the Heaviside step function, and
\begin{align} \label{eq:xa}
  x \equiv \frac1a \frac{k^2_{\rT, i} + k^2_{\rT, j}}{2k^2_{\rT,\max}} \,,
\end{align}
$k_{\rT,\max}$ being the transverse momentum of the hardest pseudo-jet at each clustering step.
The parameter $a$ in eq.~\eqref{eq:xa} is arbitrary and regulates the turn-on of the damping function.
Small non-zero values of $a$ are favourable, as the resulting clustering is closer to that of the standard anti-\kt algorithm.
Large logarithms of $a$ may however spoil the perturbative convergence if $a$ is chosen small.
The value of $a$ should therefore be tuned, possibly process-wise, to minimise unfolding effects and perturbative corrections at the same time.
Unfolding effects can be estimated by studying partonic NLO+PS predictions.
This however is beyond the scope of this study.
As suggested in Ref.~\cite{Czakon:2022wam}, we investigate the values of $a=0.05, 0.1, 0.2$.
Furthermore, there is some additional degree of freedom in the definition of $k_{\rT,\max}$.
For the \wbb process, we decided to include the transverse momentum of the final state leptons, i.e.\ $k_{\rT,\max}$ is the lepton transverse momentum if the latter is the largest transverse momentum at a given clustering step.

\begin{figure}[!t]
  \centering
  \includegraphics[width=0.89\textwidth,trim={0 15 0 10},clip]{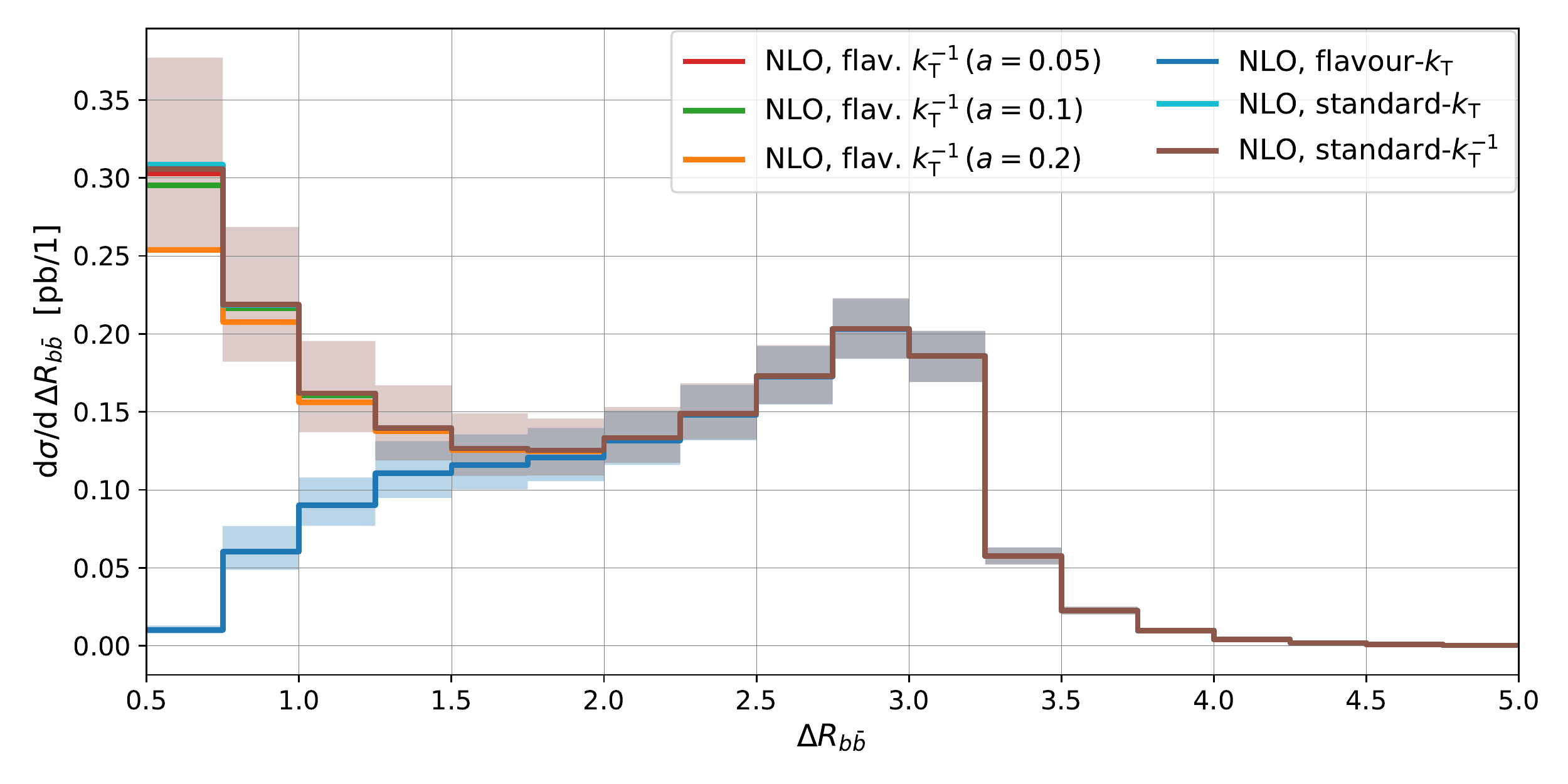}
  \caption{
    Distribution of $\Delta R_{b\bar{b}}$ at NLO for $\PW^+b\bar b$, calculated with different jet algorithms: flavour-\kt, standard \kt and anti-\kt, and flavoured anti-\kt with a selected set of values for the tuneable parameter $a$.
    The coloured bands show the scale uncertainty using the standard 7-point variation scheme for the calculations based on the flavour-\kt and standard anti-\kt algorithms.
    All calculations were performed simultaneously using the same Monte-Carlo seed.
  }
  \label{fig:jetalgos_all_dRbb}
\end{figure}

It is instructive to apply the standard \kt and anti-\kt jet algorithms and their flavour-sensitive versions at NLO, where they are all IR-safe and can be compared to each other.
In particular, let us consider the distribution of the distance between the $b$-flavoured jets, as displayed in Figure~\ref{fig:jetalgos_all_dRbb}, for $\PW^+\bbbar$ production at NLO QCD.
One can see that the distribution, obtained with the flavoured \kt algorithm, is significantly suppressed with respect to all the others in the region of small distances $\Delta R_{b\bar{b}}<1.5$.
We also note that the flavoured \kt algorithm deviates substantially even from the standard \kt prescription, which is instead aligned with all the other anti-\kt-based algorithms.
The suppression for small $\Delta R_{b\bar{b}}$ can be attributed to the modification of the beam distance in the case of the flavour \kt algorithm \cite{Banfi:2006hf}.
The reason is that, in the flavour \kt algorithm, the flavoured pseudo-jets have a larger beam distance than in the standard prescription, which favours the recombination of the $b$ and $\bar{b}$ quarks when they are close in angular separation.
We will encounter this sharp difference in the low $\Delta R_{b\bar{b}}$ region again in the next section, where we will extend the analysis to NNLO, using exclusively the flavour-sensitive jet algorithms.

\section{Phenomenology}
\label{sec:phenomenology}

In this section we discuss the results obtained with both the flavoured \kt and flavoured anti-\kt jet algorithms for $\PWp(\to\ell\nu)b\bar{b}$ production, focussing on the \PWp signature.
The previous study of the flavoured version of the anti-\kt algorithm~\cite{Czakon:2022wam} investigated examples of $\PZ+b$ and $t\bar t$ processes.
By comparing NLO+PS and NNLO QCD corrections, it was concluded that an optimal value for the jet-algorithm parameter $a$ is around $a = 0.1$.
This choice minimises the estimated unfolding corrections and, at the same time, leads to perturbatively stable predictions.

We extend the previous studies of this algorithm to a high multiplicity process: $\wbb$ production.
We structure our presentation to highlight two important aspects in particular:
\begin{itemize}
\item the difference between the flavoured \kt and flavoured anti-\kt algorithms;
\item the impact of the jet-algorithm parameter $a$.
\end{itemize}
We quantify the latter as the difference between the results for different values of $a$, which we choose as $a=0.05,0.1,0.2$ following Ref.~\cite{Czakon:2022wam}.
We therefore present differential distributions for which the difference between \kt and anti-\kt is particularly strong, to highlight its origins.
Similarly, we present differential distributions which are particularly dependent on the value of $a$, and identify the corresponding sensitive regions.

In addition, the usage of the flavoured anti-\kt algorithm puts us in a good position to compare our theoretical predictions against the experimental analyses, which are typically done using the anti-\kt algorithm.
To this end, we compare our theoretical predictions for the total cross section to the measurement of Ref.~\cite{CMS:2016eha}, showing good agreement and corroborating the need for NNLO QCD corrections in this process.

We present results for the total cross sections in Section~\ref{sec:TotCrossSection}, and for a number of differential distributions in Section~\ref{sec:DiffDistr}.
Unless explicitly indicated by `standard', we will be referring to the flavoured versions of the jet algorithms when mentioning the \kt and anti-\kt algorithms.


\subsection{Total cross sections and comparison with the CMS data}
\label{sec:TotCrossSection}

In this section we present our results for the total cross section in both the inclusive and exclusive setups.
We compare the latter against the available CMS data~\cite{CMS:2016eha}. The NLO and NNLO $K$-factors are defined by
\begin{equation}
K_\nlo  = \frac{\sigma_\nlo}{\sigma_\lo}, \qquad \qquad \qquad
K_\nnlo = \frac{\sigma_\nnlo}{\sigma_\nlo}.
\label{eq:Kfactor}
\end{equation}

\begin{table}[t!]
\centering
Inclusive $\PWp(\to\ell^+\nu) b \bar{b}$ cross sections \\
{\small
\renewcommand{\arraystretch}{1.3}
\begin{tabularx}{\textwidth}{|C{1.2}|C{1.2}|C{1.2}|C{.6}|C{1.2}|C{.6}|}
\hline
 \rowcolor{blue!9}
    Jet algorithm
  & $\sigma_\text{LO}$ [fb]
  & $\sigma_\text{NLO}$ [fb]
  & $K_\text{NLO}$
  & $\sigma_\text{NNLO}$ [fb]
  & $K_\text{NNLO}$
  \\
\hline
    flavour-$\kt$
      & $213.24(8)^{+21.4\%}_{-16.1\%}$
      & $362.0(6)^{+13.7\%}_{-11.4\%}$
      & 1.70
      & $445(5)^{+6.7\%}_{-7.0\%}$
      & 1.23
      \\
\hline
      flavour anti-$\kt$ ($a=0.05$)
      & $262.52(10)^{+21.4\%}_{-16.1\%}$
      & $500.9(8)^{+16.1\%}_{-12.8\%}$
      & 1.91
      & $690(7)^{+10.9\%}_{-9.7\%}$
      & 1.38
      \\
\hline
      flavour anti-$\kt$ ($a=0.1$)
      & $262.47(10)^{+21.4\%}_{-16.1\%}$
      & $497.8(8)^{+16.0\%}_{-12.7\%}$
      & 1.90
      & $677(7)^{+10.4\%}_{-9.4\%}$
      & 1.36
      \\
  \hline
      flavour anti-$\kt$ ($a=0.2$)
      & $261.71(10)^{+21.4\%}_{-16.1\%}$
      & $486.3(8)^{+15.5\%}_{-12.5\%}$
      & 1.86
      & $647(7)^{+9.5\%}_{-8.9\%}$
      & 1.33
      \\
\hline
\end{tabularx}
\caption{\label{tab:xsection}
  Fiducial cross sections for \emph{inclusive} $pp\to \PWp(\to\ell^+\nu) b \bar{b}$ production at the LHC with $\sqrt{s}~=~8$~TeV at LO, NLO and NNLO QCD using the flavour-\kt and flavour anti-\kt algorithms.
The corresponding $K$-factors are defined in Eq.~\ref{eq:Kfactor}.
The statistical errors are shown in parentheses and correspond to the central predictions, while sub- and superscripts denote the theoretical uncertainty calculated using the standard 7-point scale variation.
}
}
\end{table}

\begin{table}[t!]
\centering
Exclusive $\PW^\pm(\to\ell^\pm\nu) b \bar{b}$ cross sections \\
{\small
\renewcommand{\arraystretch}{1.3}
\begin{tabularx}{\textwidth}{|C{1.1}|C{1.2}|C{1.35}|C{.55}|C{1.3}|C{.5}|}
\hline
 \rowcolor{blue!9}
    Jet algorithm
  & $\sigma_\text{LO}$ [fb]
  & $\sigma_\text{NLO}$ [fb]
  & $K_\text{NLO}$
  & $\sigma_\text{NNLO}$ [fb]
  & $K_\text{NNLO}$
  \\
\hline
        flavour-\kt
        & $345.97(9)^{+21.4\%}_{-16.2\%}$
        & $408.4(5)^{+4.2\%}_{-6.2\%}\,^{(41\%)}_{(28\%)}$
        & 1.18
        & $434(8)^{+1.7\%}_{-2.5\%}\,^{(16\%)}_{(16\%)}$
        & 1.06
        \\
\hline
        flavour anti-\kt ($a=0.05$)
        & $425.71(12)^{+21.5\%}_{-16.2\%}$
        & $540.3(7)^{+6.2\%}_{-7.4\%}\,^{(42\%)}_{(29\%)}$
        & 1.27
        & $636(11)^{+5.4\%}_{-5.0\%}\,^{(23\%)}_{(20\%)}$
        & 1.18
        \\
\hline
        flavour anti-\kt ($a=0.1$)
        & $425.63(12)^{+21.5\%}_{-16.2\%}$
        & $538.7(7)^{+6.1\%}_{-7.4\%}\,^{(42\%)}_{(29\%)}$
        & 1.27
        & $630(10)^{+5.0\%}_{-4.8\%}\,^{(22\%)}_{(20\%)}$
        & 1.17
        \\
\hline
        flavour anti-\kt ($a=0.2$)
        & $424.37(12)^{+21.5\%}_{-16.2\%}$
        & $530.6(7)^{+5.8\%}_{-7.2\%}\,^{(42\%)}_{(29\%)}$
        & 1.25
        & $606(10)^{+4.2\%}_{-4.2\%}\,^{(21\%)}_{(19\%)}$
        & 1.14
        \\
\hline
\end{tabularx}

\caption{\label{tab:xsection_exc}
Fiducial cross sections for \emph{exclusive} $pp\to \PW(\to\ell\nu) b \bar{b}$ (combined W$^\pm$ signature) production at the LHC with $\sqrt{s}~=~8$~TeV at LO, NLO and NNLO QCD using the flavour-\kt and flavour anti-\kt algorithms.
The structure is the same as in Table~\ref{tab:xsection}, except we additionally provide the uncorrelated scale variation (shown in parentheses).
}
}
\end{table}

The integrated cross-sections for the \PWp signature in the inclusive setup are presented in Table~\ref{tab:xsection}.
We also report the theoretical uncertainty of the cross section due to missing higher orders, which we estimate as customary by varying the renormalisation and factorisation scales.
The scale dependence of the inclusive cross section is estimated using the standard 7-point scale variation, where the renormalisation $\mu_R$ and factorisation $\mu_F$ scales are varied according to
\begin{equation}
\bigg(\frac{\mu_R}{H_\RT},\frac{\mu_F}{H_\RT}\bigg) \in \big\lbrace
(0.5,0.5),(1,0.5),
(0.5,1),(1,1),(2,1),
(1,2),(2,2)
\big\rbrace,
\end{equation}
where $H_\RT$ is the chosen central scale, defined in Eq.~\eqref{eq:HTdef}.

The results for the flavour-$\kt$ algorithm were already presented in Ref.~\cite{Hartanto:2022qhh}, and are provided here for comparison.
The relative difference between the flavoured algorithms is significant, reaching $\sim 50\%$ at NNLO, and gets modified with perturbative corrections due to different $K$-factors.
As pointed out in Section~\ref{sec:calculation}, this difference comes from the particular behaviour of the flavour modification of the \kt algorithm in the small $\Delta R_{b\bar{b}}$ region.
The scale uncertainty of the anti-\kt setups is larger than the uncertainty of the flavour-\kt algorithm and depends only slightly on the jet-algorithm parameter $a$.
The $K$-factors are also larger by the same amount, and increase for decreasing $a$-parameter, indicating an expected breakdown of the perturbative convergence due to logarithms of small $a$.
In the next section we will see that the larger values and scale uncertainties of the anti-\kt setup originate from the region where the distance between the $b$-flavoured jets $\Delta R_{\bbbar}$ is small.

\begin{figure}[!t]
  \centering
  \includegraphics[width=0.89\textwidth,trim={0 0 0 30},clip]{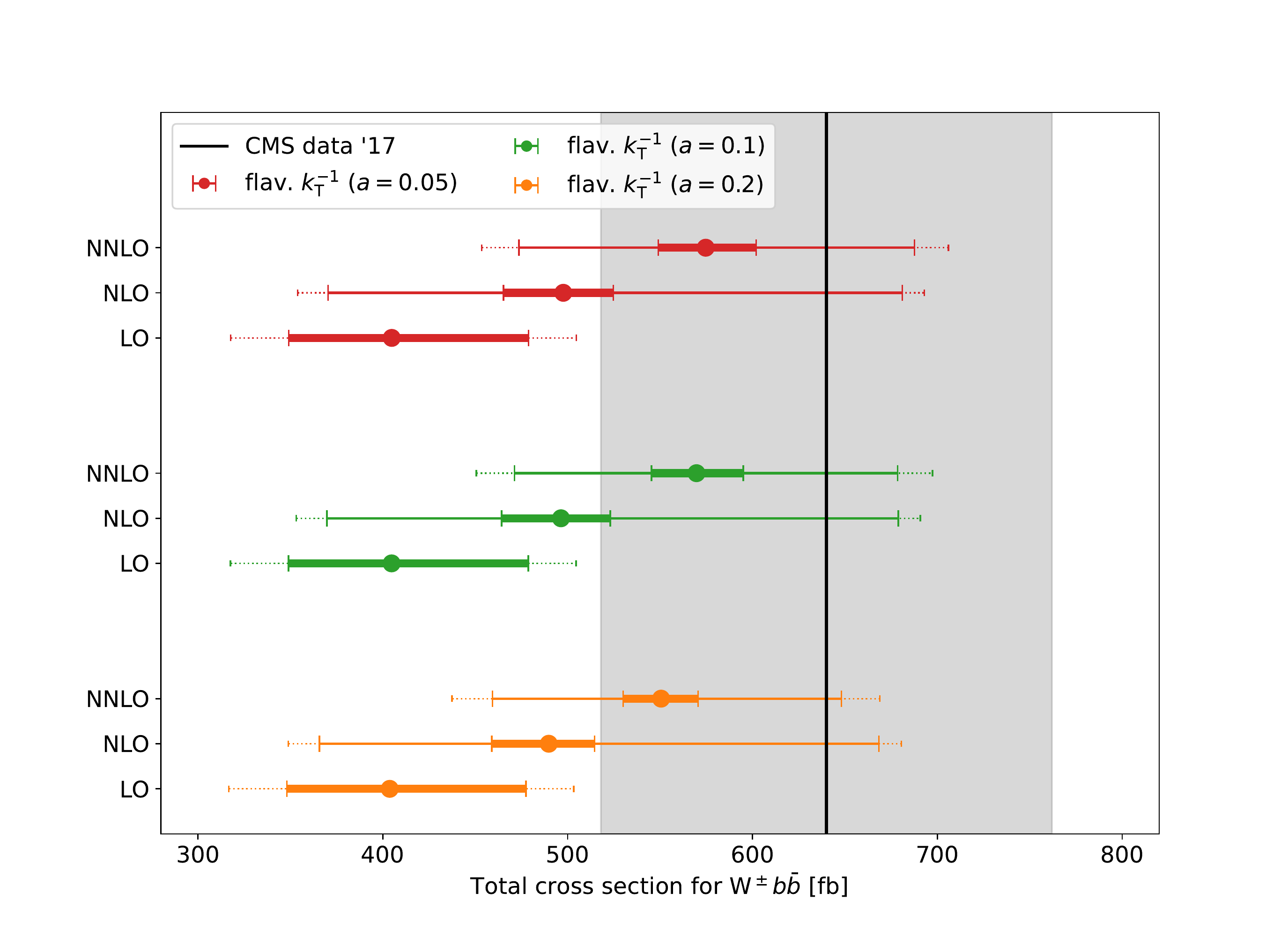}
  \caption{
    Comparison between CMS data from Ref.~\cite{CMS:2016eha} and the theoretical predictions using the flavoured anti-\kt jet algorithm with different $a$-parameters in the exclusive setup for the \PWp and \PWm combined signature.
    The theoretical uncertainty is estimated by the standard 7-point scale variation (thick band).
    At NLO and NNLO we also show the theoretical uncertainty calculated using the uncorrelated prescription as described in the text (thin band).
    The multiplicative hadronisation and additive DPI correction factors are taken into account in all theoretical predictions.
    Additionally, we include (in quadrature) uncertainties on the DPI factor and hadronisation corrections \cite{CMS:2016eha} to the uncorrelated theoretical uncertainties via a dotted extension to the bands.
  }
  \label{fig:CMS-xsecs}
\end{figure}

Next, we turn our attention to the exclusive setup.
The measurement of Ref.~\cite{CMS:2016eha} presented a cross section in the exclusive setup for the combined $\PW^\pm$ signature.
The corresponding theoretical total cross sections, obtained using both the flavoured \kt and anti-\kt jet algorithms, are presented in Table~\ref{tab:xsection_exc}.
As in the inclusive case, the theoretical uncertainty is estimated using the standard 7-point scale variation.
In addition, in order to mitigate the well-known underestimation of theoretical uncertainty in the exclusive setup, we also include the more conservative uncorrelated scale variation, proposed in Ref.~\cite{Stewart:2011cf}.
To obtain the uncorrelated theoretical uncertainty, we write the exclusive \wbb cross section at NLO (NNLO) as
\begin{align}
  \sigma_{\text{(N)}\nlo,\wbb,\exc}  & = \sigma_{\text{(N)}\nlo,\wbb,\inc} - \sigma_{\text{(N)}\lo,\wbbj,\inc}\,.
\end{align}
The first term on the right-hand side is the original inclusive \wbb production, while the second term is the inclusive \wbb production in association with a hard jet.
The theoretical uncertainty of the exclusive \wbb cross section is then taken to be
\begin{align}
  \Delta\sigma_{\text{(N)}\nlo,\wbb,\exc}  &= \sqrt{\left(\Delta\sigma_{\text{(N)}\nlo,\wbb,\inc}\right)^2  + \left(\Delta\sigma_{\text{(N)}\lo,\wbbj,\inc}\right)^2},
\end{align}
where $\Delta\sigma_{\text{(N)}\nlo,\wbb,\inc}$ and $\Delta\sigma_{\text{(N)}\lo,\wbbj,\inc}$ are the theoretical uncertainties of the \wbb and \wbbj inclusive cross sections respectively, as obtained from direct 7-point scale variation.
We observe that the corrections, as well as the standard and uncorrelated scale-variation uncertainties, behave similarly to the inclusive case.

In Figure~\ref{fig:CMS-xsecs} we show the comparison of the CMS measurement against our theoretical predictions, obtained with the flavoured anti-\kt algorithm.
We note that our theoretical predictions include both the multiplicative hadronisation correction factor ($0.81 \pm 0.07$ pb) and the additive double parton interaction (DPI) correction ($0.06 \pm 0.06$ pb), as specified in Ref.~\cite{CMS:2016eha}.
Since uncertainties on these factors do not originate from fixed order predictions, we separate them from the estimate of missing higher orders.
After being added in quadrature to the uncorrelated theoretical uncertainties, 
they contribute to the full theoretical uncertainty, shown by dotted error bars in Figure~\ref{fig:CMS-xsecs}.
We checked that our NLO results are compatible with the NLO predictions shown in Figure~3 of Ref.~\cite{CMS:2016eha}.
The NNLO QCD corrections appear to be significant, and their inclusion in the theoretical prediction improves substantially the agreement with the experimental data.
Moreover, we find consistent agreement for all the considered values of $a$.


\subsection{Differential distributions}
\label{sec:DiffDistr}

In this section we present results for a number of differential distributions, focusing on the inclusive $\PWp(\to\ell^+\nu)b\bar{b}$ production.
We selected distributions which showcase the difference between the flavour \kt and anti-\kt algorithms or exhibit a particularly strong dependence on the $a$-parameter.

\begin{figure}[!t]
  \centering
  \includegraphics[width=0.49\textwidth,trim={0 20 50 60},clip]{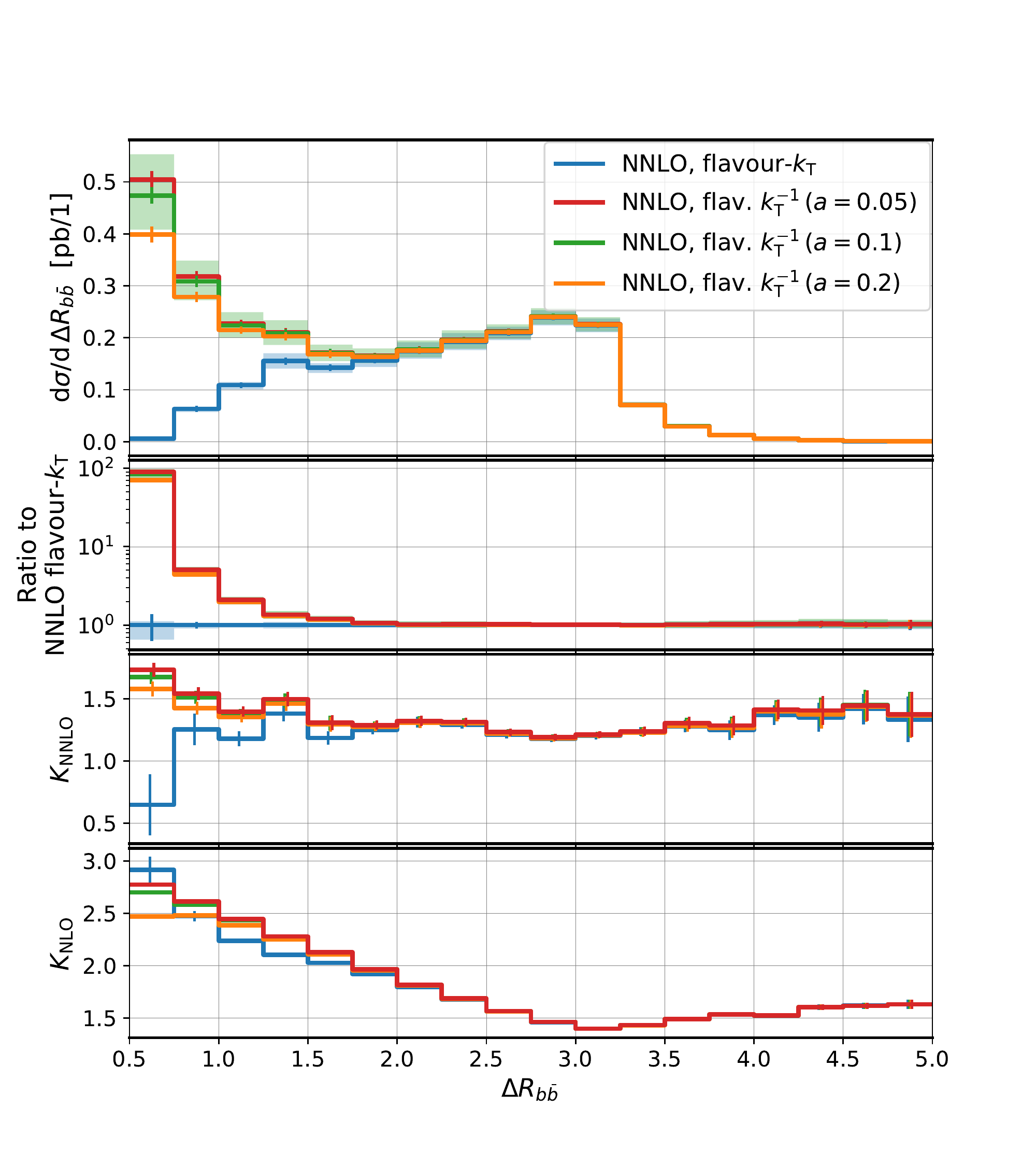}
  \hfill
  \includegraphics[width=0.49\textwidth,trim={0 20 50 60},clip]{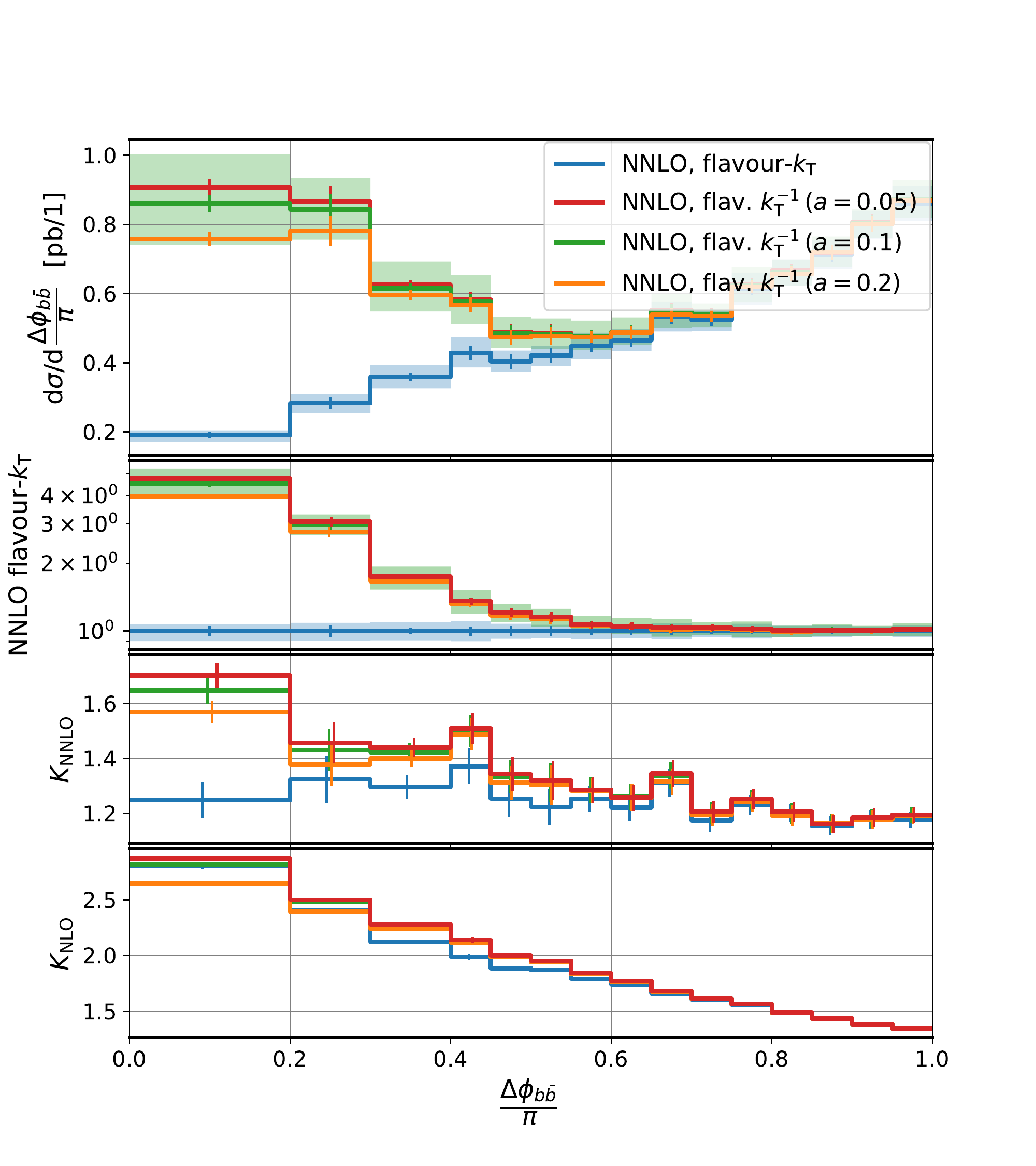}
  \caption{
    Distribution of $\Delta R_{b\bar{b}}$ and $\Delta \phi_{b\bar{b}}$ for inclusive $pp\to\PW^+(\to\ell^+\nu)b\bar{b}$ production.
    The second panel shows the ratio of all setups to the flavoured-\kt algorithm.
    The coloured bands define scale uncertainty for two calculations: flavour-\kt, and flavoured anti-\kt with $a=0.1$.
    The last two panels show the $K$-factors at NNLO and NLO, correspondingly.
    The vertical bars define the statistical uncertainty.
    All calculations were performed simultaneously using the same Monte-Carlo seed.
  }
  \label{fig:jetalgos_dRbb_dphibb}
\end{figure}

We start by showing the distributions for the distance between the $b$-flavoured jets $\Delta R_{\bbbar}$ and their azimuthal angular separation $\Delta\phi_{\bbbar}$ in Figure~\ref{fig:jetalgos_dRbb_dphibb}.
These distributions are of particular interest, as they explicitly enter the definition of the jet algorithms.
One can clearly see that the region of small angular separation is the origin of the increased cross section and scale uncertainty of the anti-\kt setups in comparison to the \kt setup.
The reason for this enhancement, which can be observed already at NLO, is the sensitivity to gluon splittings into $b$-quark pairs.
These splittings are divergent, and are regulated by the jet definitions and phase space cuts.
The flavoured \kt suppresses this region and therefore is less sensitive to the gluon splittings.
In both distributions the differences between the \kt and anti-\kt algorithms become slightly stronger in this region when higher order corrections are included.
For larger angular separations, we observe that the scale uncertainty and the higher order corrections are similar for the flavour \kt and anti-\kt algorithms.
The behaviour due to this enhancement for the anti-\kt algorithms propagates to the total cross sections, discussed in Section~\ref{sec:TotCrossSection}.

As for the $a$-parameter, the results are sensitive to it at $0.5 < \Delta R_{\bbbar} < 1$.
Lower values of $\Delta R_{\bbbar}$ are not allowed due to the selected jet radius $R_{\bbbar} = 0.5$.
The differences between the selected anti-\kt setups reach 25\%. Similar observations hold for the $\Delta\phi_{\bbbar}$ distribution in the $\Delta\phi_{\bbbar}<0.4$ region.

\begin{figure}[!t]
  \centering
  \includegraphics[width=0.49\textwidth,trim={0 30 50 60},clip]{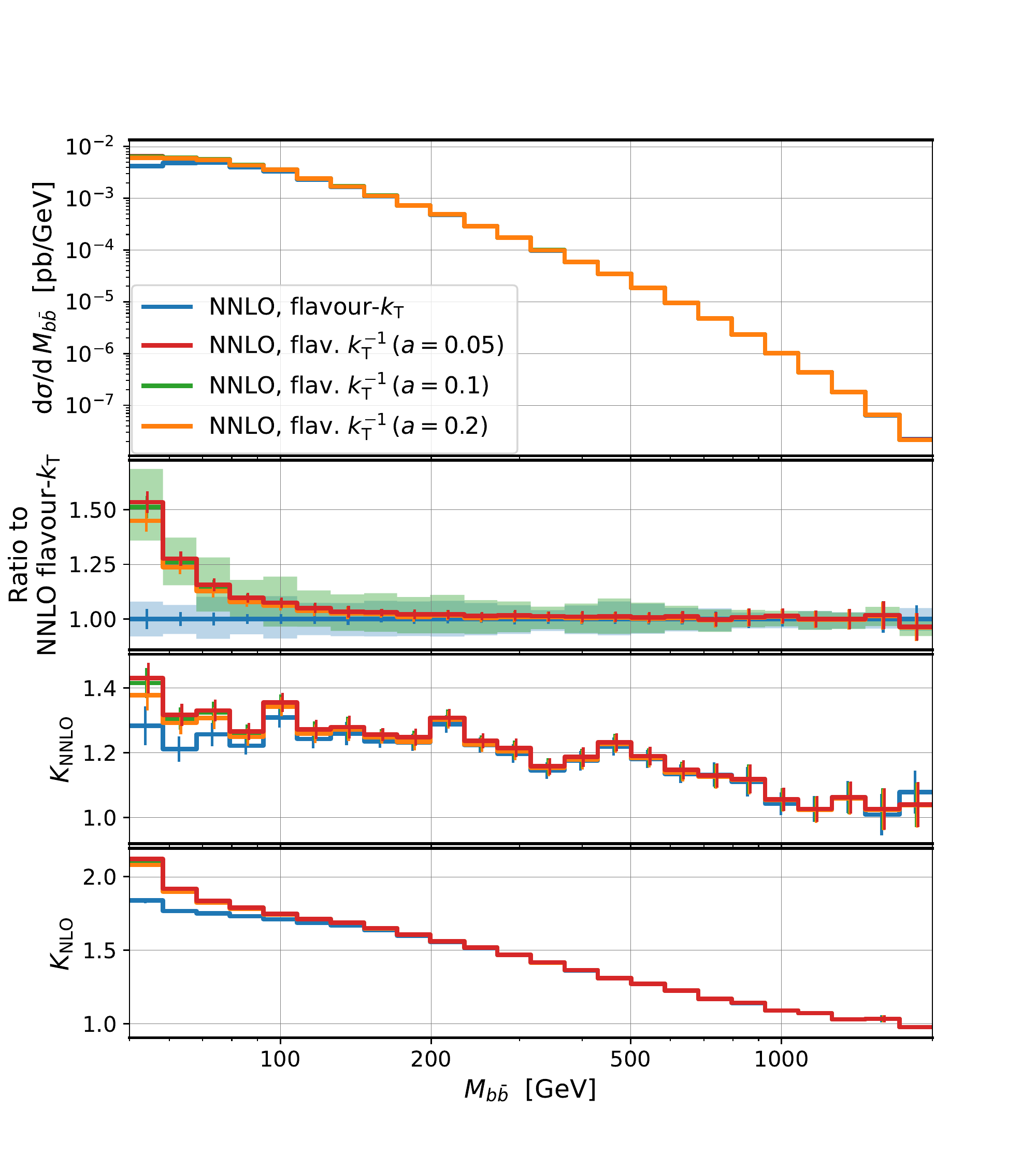}
  \hfill
  \includegraphics[width=0.49\textwidth,trim={0 30 50 60},clip]{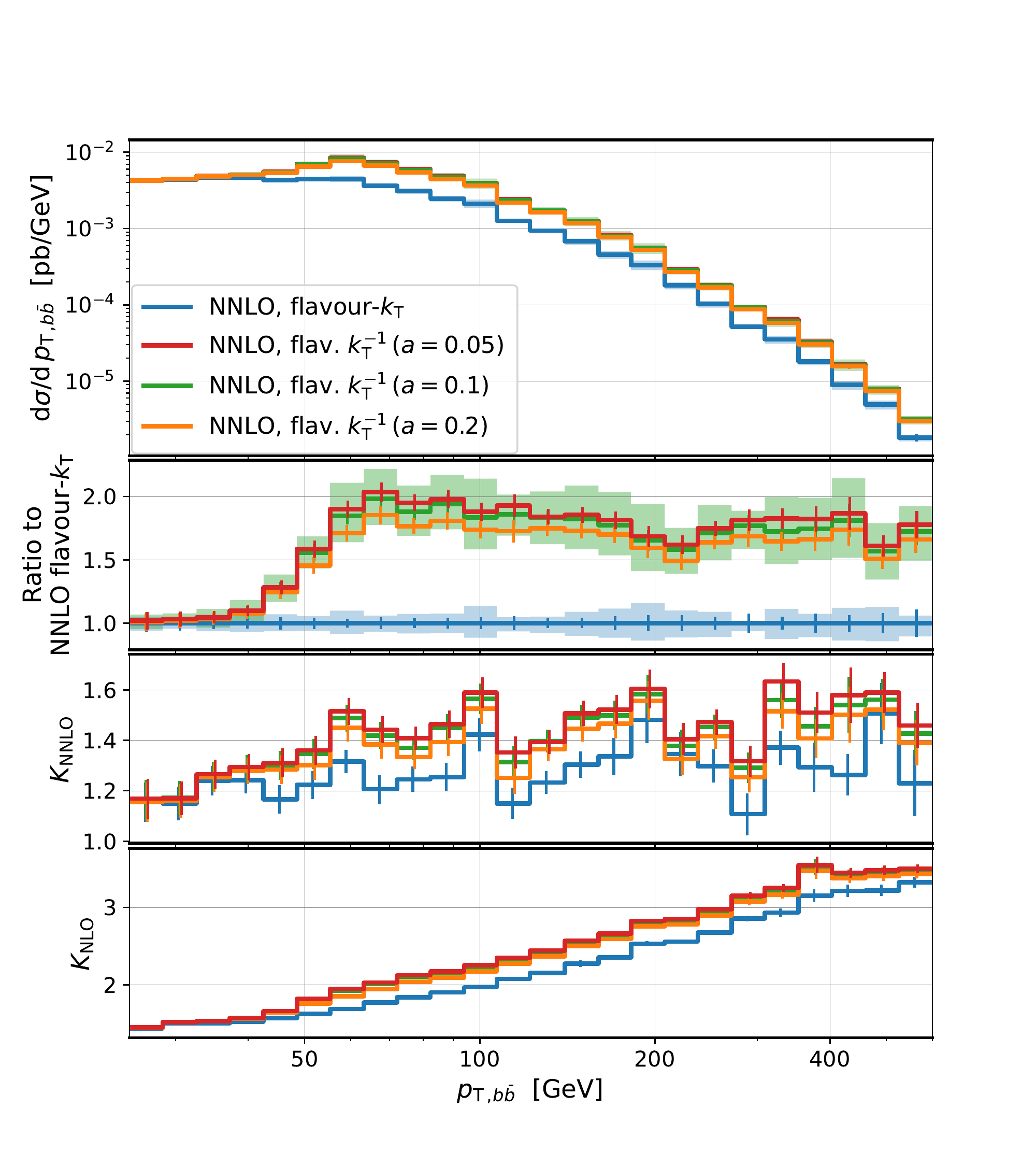}
  \caption{
    Distribution of invariant mass (left) and transverse momentum (right) of the $b\bar b$-pair for inclusive $pp\to\PW^+(\to\ell^+\nu)b\bar{b}$ production.
    The individual plot structure is the same as in Figure~\ref{fig:jetalgos_dRbb_dphibb}.
  }
  \label{fig:jetalgos_mbb_ptb1}
\end{figure}

On the left plot in Figure~\ref{fig:jetalgos_mbb_ptb1}, we present the distribution of the invariant mass of the two $b$-jets.
We observe a difference of order 50\% between the flavoured anti-\kt and \kt algorithms at low energies, starting at 50 \GeV, gradually vanishing by 200 \GeV.
Once again, we see that the scale band and results are slightly larger for the flavoured anti-\kt setups.
This feature can also be attributed to the sensitivity to gluon splittings.
With respect to the distributions in Figure~\ref{fig:jetalgos_dRbb_dphibb}, the sensitivity to the $a$-parameter at low energies is in this case smaller.
The setup at $a=0.2$ is further away from the other two, reflecting the increasing modification of the clustering sequence due to the flavour modifications.
The differences in the $\cK$-factors at NLO are completely determined by the standard jet algorithm differences, whereas at NNLO the $\cK$-factors are very similar.

The right plot of Figure~\ref{fig:jetalgos_mbb_ptb1} shows the \pt distribution of the $b$-jet pair.
The difference between \kt and anti-\kt, in this case, shows up at higher energies, starting at 40 \GeV and reaching a factor of 2.
A high \pt of the pair means that the jets are moving in the same direction.
We observe that the relative distance between the curves with different values of the $a$ parameter is roughly constant above 60 \GeV.
This might indicate that the proximity of the $b$-jets depends only weakly on the \pt of the pair.
The peak of the absolute value distribution is shifted from 35 to 60 \GeV for the anti-\kt algorithm due to the contribution from the region of small distance between the quarks in the \bbbar-pair.
This feature is observed already at NLO when using the standard anti-\kt algorithm.

\begin{figure}[!h]
  \centering
  \includegraphics[width=0.49\textwidth,trim={0 30 50 60},clip]{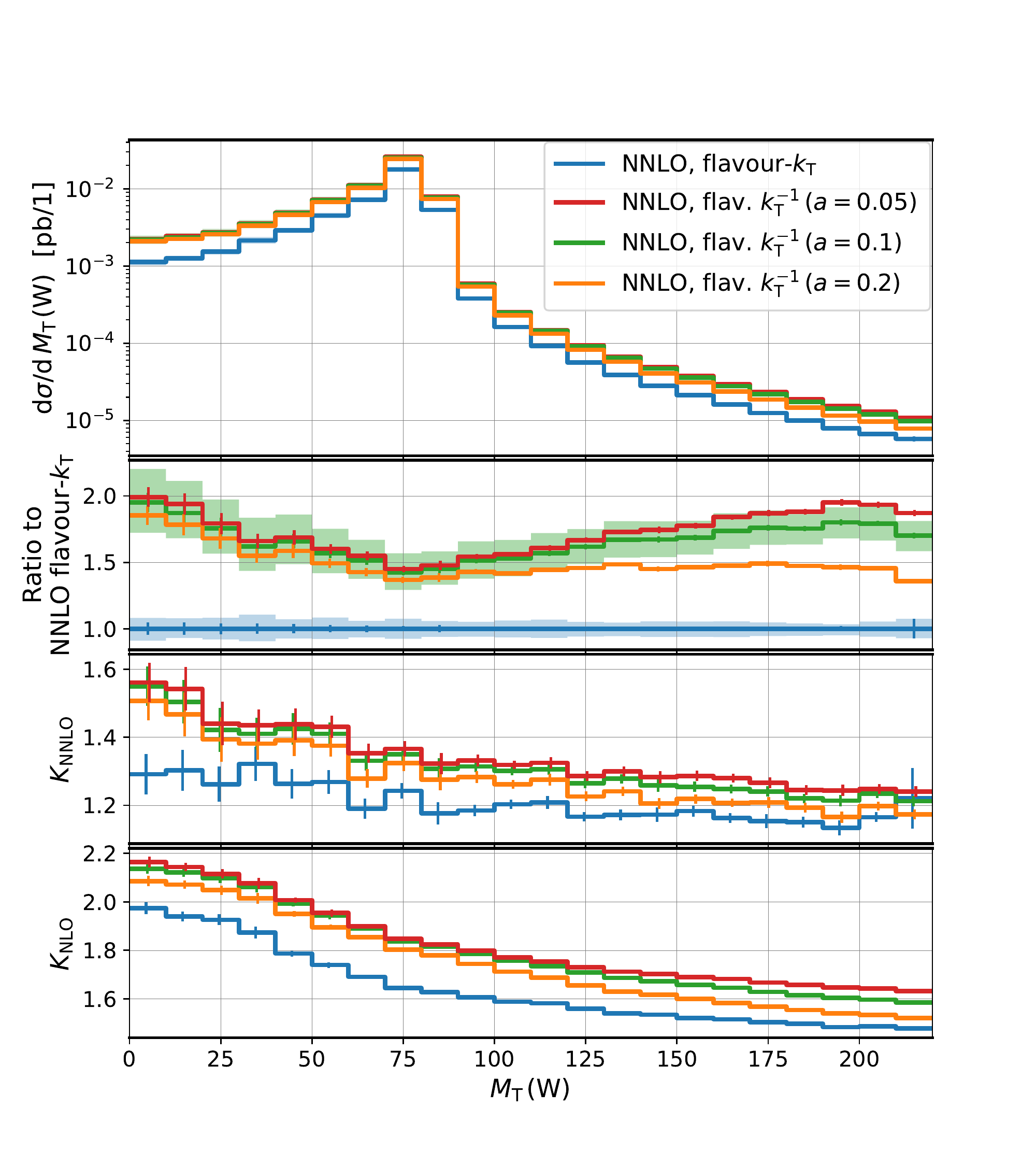}
  \hfill
  \includegraphics[width=0.49\textwidth,trim={0 30 50 60},clip]{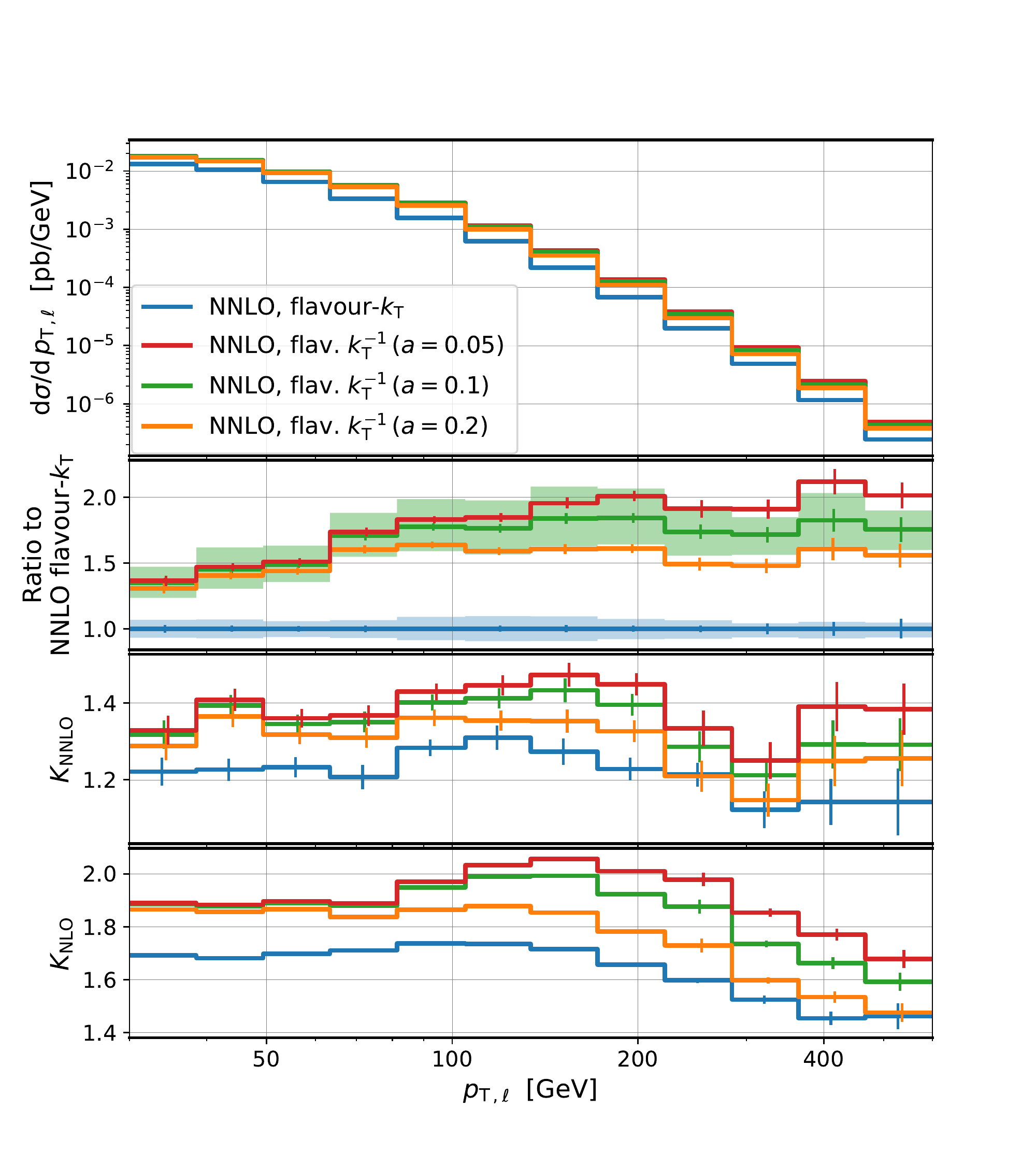}
  \caption{
  Distribution of the \PWp-boson transverse mass (left) and of the positively charged lepton transverse momentum (right)
  for inclusive $pp\to\PW^+(\to\ell^+\nu)b\bar{b}$ production.
  The individual plot structure is the same as in Figure~\ref{fig:jetalgos_dRbb_dphibb}.
  }
  \label{fig:jetalgos_mTW_pTlep}
\end{figure}

In Figure~\ref{fig:jetalgos_mTW_pTlep} we show two observables which are particularly sensitive to the $a$-parameter.
On the left, starting at 100 \GeV, the W-boson transverse mass shows a noticeable split between the anti-\kt setups with different values of $a$.
The differences between the selected setups reach 25\% and extend beyond the scale band.
For larger values of $a$, we also observed that the distribution tails of the anti-\kt setups may go even below the \kt setup.
This sensitivity is already present at LO, and it is thus not an artefact of the QCD corrections.
The region of large transverse mass is populated by events with a large recoil against the $b\bar{b}$ system, and thus events where the $b\bar{b}$ pair are kinematically close to each other.

As shown on the right of Figure~\ref{fig:jetalgos_mTW_pTlep}, the charged lepton \pt distribution also shows significant sensitivity to the $a$-parameter, starting at 60 \GeV and reaching 25\% by 300 \GeV.
We note, however, that this sensitivity is not observed at LO, where the setups match each other almost completely, and only starts occurring at NLO, as is reflected by the corresponding $K$-factor.
The NNLO corrections amplify these differences.

\section{NNLO versus NLO-merged calculations}
\label{sec:NLOplus}

The large corrections and strong dependence on the renormalisation and factorisation scales in the NLO QCD calculation of the inclusive \wbb production originate from the tree-level $qg(\bar{q}g)\to \wbb q (\bar{q})$ subprocess, which opens up only at
this order~\cite{Ellis:1998fv,FebresCordero:2006nvf,FebresCordero:2009xzo,Badger:2010mg}.
Efforts have been made to evaluate such subprocesses at higher accuracy by computing the $pp \to\wbbj$ process at 
NLO QCD~\cite{Reina:2011mb,Luisoni:2015mpa}.
$pp \to\wbbj$ production at NLO QCD constitutes one of the ingredients of NNLO QCD calculation of \wbb production.
The NLO QCD prediction for \wbb production can then be improved by taking into account the contribution from \wbbj production computed also at NLO QCD accuracy, achieved through a particular merging scheme.
Such an NLO-merged prediction contains some of the NNLO QCD corrections.
The aim of this section is to compare the NNLO prediction for inclusive \wbb production
against the NLO-merged calculation and assess the genuine impact of the full NNLO QCD corrections.

We employ only fixed order calculations to carry out the comparisons and focus on the $\wpbb$ signature.
To this end, we make use of the \textit{exclusive sums} method to merge fixed order
NLO QCD calculations with different jet multiplicities~\cite{SM:2012sed,Anger:2017glm}.
The resulting NLO-merged prediction will further be denoted as `NLO+'.
The inclusive NLO+ prediction for $\wbb$ production is derived by summing up the exclusive NLO QCD $\wbb$
and inclusive NLO QCD $\wbbj$ contributions
\begin{equation}
\sigma_{\nloplus,\wbb,\inc} = \sigma_{\nlo,\wbb,\exc} + \sigma_{\nlo,\wbbj,\inc}.
\label{eq:nloplus}
\end{equation}
Note that, if the last term of Eq.~\eqref{eq:nloplus} were evaluated at leading order, we would recover the NLO inclusive cross section for $\wbb$ production.

\begin{table}[t]
\centering
Inclusive $\PWp(\to\ell^+\nu) b \bar{b}$ cross sections \\
{\small
\renewcommand{\arraystretch}{1.3}
\begin{tabularx}{\textwidth}{|C{1.3}|C{1.2}|C{1.2}|C{.5}|C{1.2}|C{.6}|}
\hline
 \rowcolor{blue!9}
    Jet algorithm
  & $\sigma_\text{NLO}$ [fb]
  & $\sigma_\text{NNLO}$ [fb]
  & $K_\text{NNLO}$
  & $\sigma_\text{NLO+}$ [fb]
  & $K_\text{NLO+}$
  \\
\hline
    flavour-$\kt$
      & $362.0(6)^{+13.7\%}_{-11.4\%}$
      & $445(5)^{+6.7\%}_{-7.0\%}$
      & 1.23
      & $426(5)^{+7.6\%}_{-8.9\%}$
      & 1.18
      \\
\hline
     flavour anti-$\kt$ ($a=0.05$)
      & $500.9(8)^{+16.1\%}_{-12.8\%}$
      & $690(7)^{+10.9\%}_{-9.7\%}$
      & 1.38
      & $635(6)^{+11.2\%}_{-11.1\%}$
      & 1.27
      \\
\hline
      flavour anti-$\kt$ ($a=0.1$)
      & $497.8(8)^{+16.0\%}_{-12.7\%}$
      & $677(7)^{+10.4\%}_{-9.4\%}$
      & 1.36
      & $626(6)^{+10.8\%}_{-10.9\%}$
      & 1.26
      \\
  \hline
      flavour anti-$\kt$ ($a=0.2$)
      & $486.3(8)^{+15.5\%}_{-12.5\%}$
      & $647(7)^{+9.5\%}_{-8.9\%}$
      & 1.33
      & $602(6)^{+10.2\%}_{-10.5\%}$
      & 1.24
      \\
\hline
\end{tabularx}
\caption{
\label{tab:xsectionNLO+}
NLO, NNLO and NLO+ fiducial cross sections for \emph{inclusive} $pp\to \PWp(\to\ell^+\nu) b \bar{b}$ production at the LHC with $\sqrt{s}~=~8$~TeV
using the flavour-\kt and flavour anti-\kt algorithms.
The corresponding $K$-factors are defined in Eq.~\ref{eq:Kfactor2}.
The statistical errors are shown in parentheses and correspond to the central predictions, while sub- and superscripts denote the theoretical uncertainty calculated using the standard 7-point scale variation.
}
}
\end{table}

We first present in Table~\ref{tab:xsectionNLO+} the comparison between NNLO and NLO+ predictions at the level of fiducial cross section.
The NLO, NNLO and NLO+ cross sections are presented together with the corresponding $K$-factors defined by
\begin{equation}
K_\nnlo    = \frac{\sigma_\nnlo}{\sigma_\nlo}, \qquad \qquad \qquad
K_\nloplus = \frac{\sigma_\nloplus}{\sigma_\nlo}.
\label{eq:Kfactor2}
\end{equation}
We note that the NLO+ predictions capture already a significant amount of the NNLO corrections, although they still come short in predicting the NNLO cross sections, about 4\% lower for the flavour-$k_\text{T}$ prediction, and up to 8\% lower for the flavour anti-$k_\text{T}$ predictions.
It is important to note that the NLO+ predictions already exhibit a much improved scale dependence over NLO results, basically resembling NNLO level.

\begin{figure}[t!]
 \begin{center}
   \includegraphics[width=0.485\textwidth]{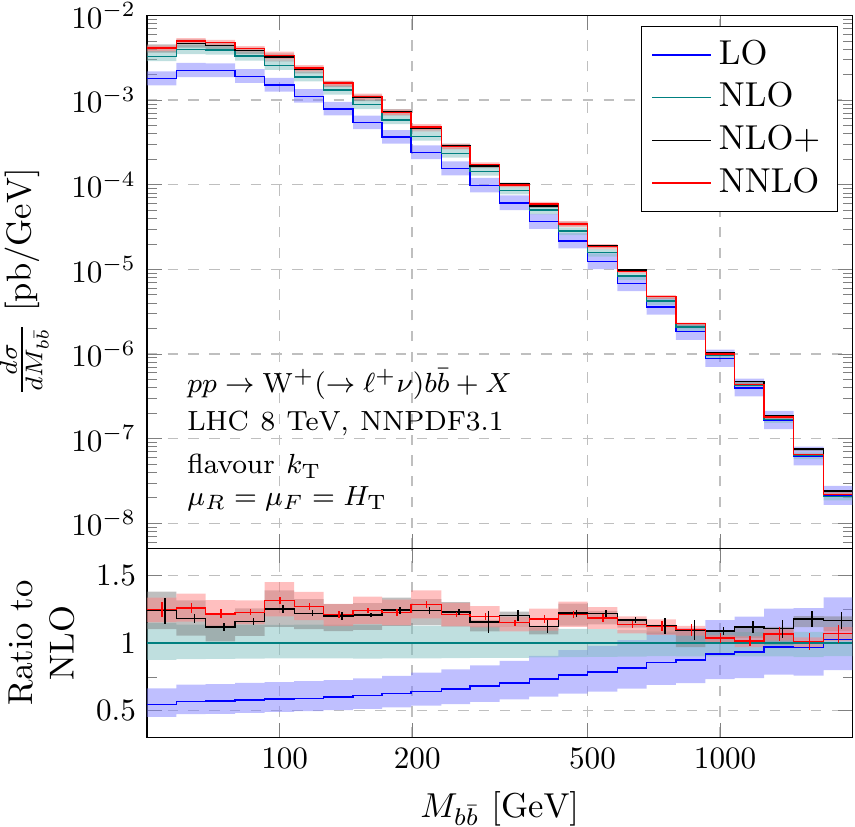}\;\;
   \includegraphics[width=0.485\textwidth]{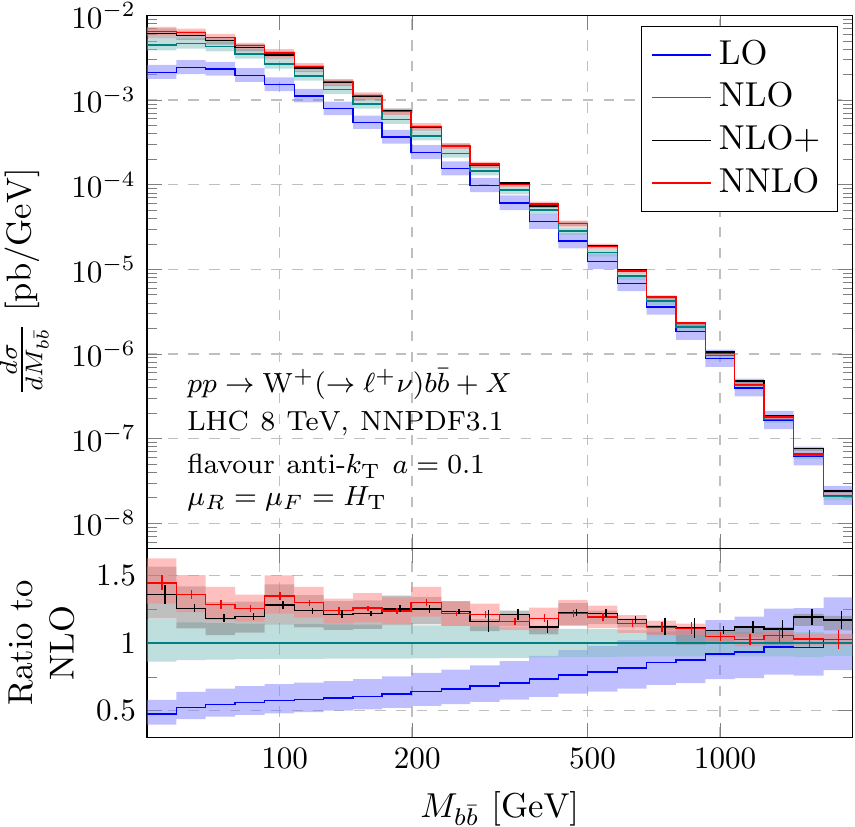}%
   \caption{$M_{b\bar{b}}$ differential distributions for inclusive $pp\to \PWp(\to\ell^+\nu) b \bar{b}$ production obtained using
   the flavour-$k_\text{T}$ (left) and flavour anti-$k_\text{T}$ (right) jet algorithms. LO, NLO, NNLO and NLO+ predictions are presented.
   The lower panel shows the ratio to the NLO QCD calculation.
   The vertical bars define the statistical uncertainty.
   }
 \label{fig:MbbNLOplus}
 \end{center}
\end{figure}

\begin{figure}[t!]
 \begin{center}
   \includegraphics[width=0.485\textwidth]{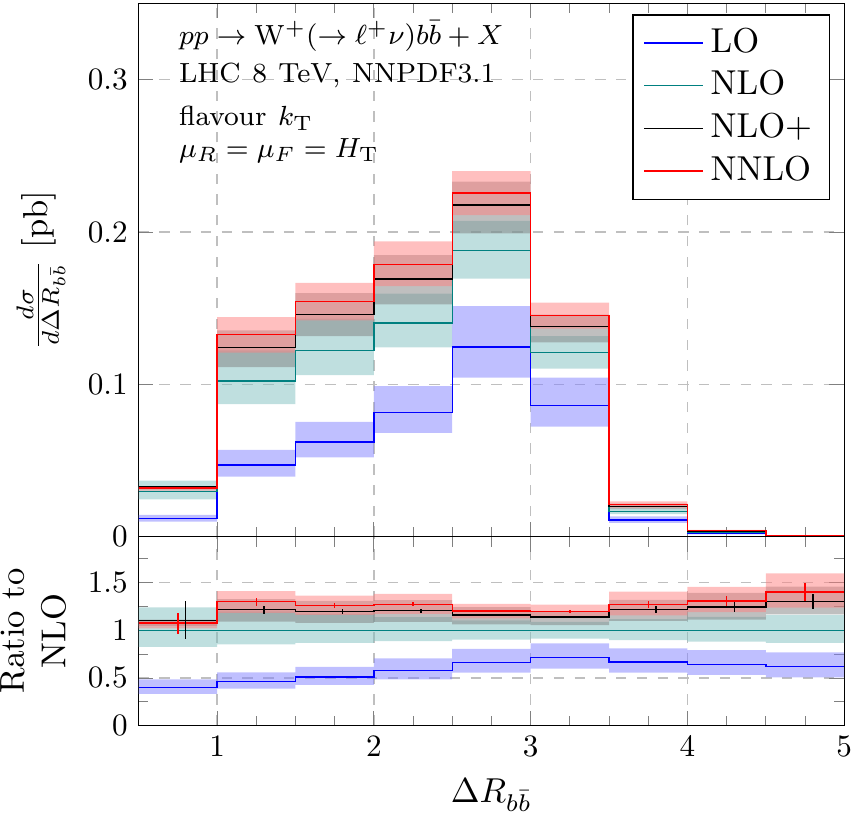}\;\;
   \includegraphics[width=0.485\textwidth]{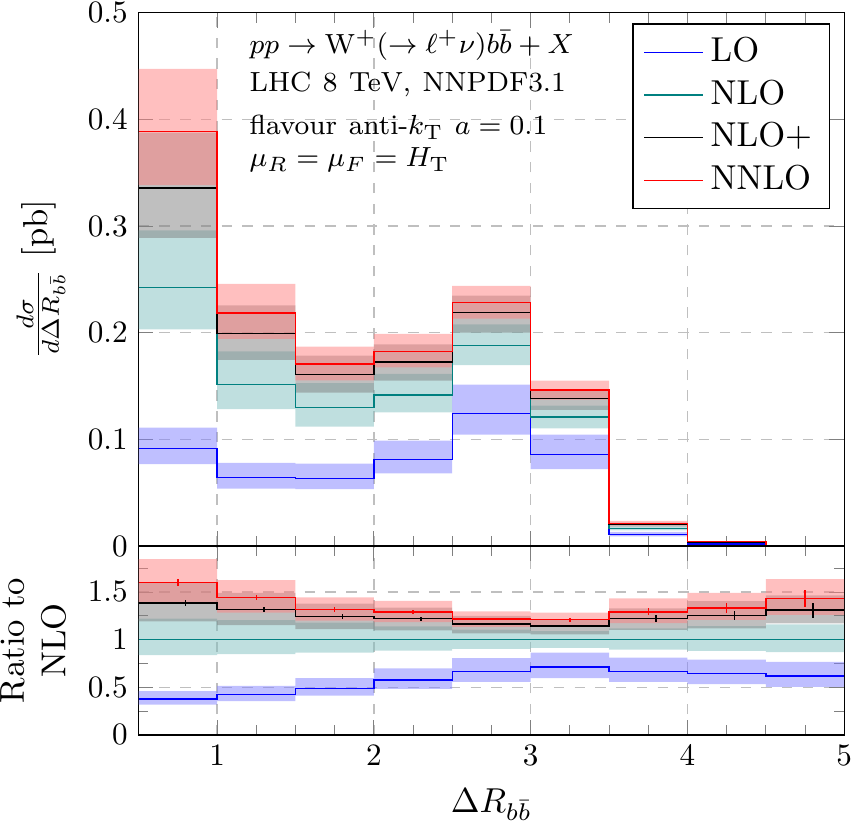}%
   \caption{$\Delta{R}_{b\bar{b}}$ differential distributions for inclusive $pp\to \PWp(\to\ell^+\nu) b \bar{b}$ production obtained using
   the flavour-$k_\text{T}$ (left) and flavour anti-$k_\text{T}$ (right) jet algorithms.
   The individual plot structure is the same as in Figure~\ref{fig:MbbNLOplus}.
   }
   \label{fig:dRbbNLOplus}
 \end{center}
\end{figure}

Comparisons between the diffential distributions for NNLO and NLO-merged calculations are shown in Figures~\ref{fig:MbbNLOplus}~and~\ref{fig:dRbbNLOplus}.
In particular, we show the invariant mass of the $b\bar{b}$ system ($M_{b\bar{b}}$), and 
the separation in the rapidity and azimuthal angle between the two leading $b$-jets ($\Delta R_{b\bar{b}}$).
We display the LO, NLO, NNLO and NLO+ predictions obtained using the flavour-$k_\text{T}$ jet algorithm (left plots) and the flavour anti-$k_\text{T}$ jet algorithm with $a=0.1$ (right plots).
In general, we also observe the same characteristics, already highlighted in the comparison of the fiducial cross sections.
Namely, that the behaviour and scale dependence of NLO+ predictions is markedly similar to NNLO QCD corrections.
The NLO+ results generally exhibit a lower normalisation as already observed in the fiducial cross section comparison. 
For the differential distributions, however, in some phase-space regions, NLO+ calculations display higher predictions than the NNLO ones
which can be particularly seen in the tails of $M_{b\bar{b}}$ distribution (c.f.\ Figure~\ref{fig:MbbNLOplus}).

\section{Conclusions}
\label{sec:conclusions}

We considered the production and leptonic decay of a $\PW$-boson in association with a bottom-quark pair at the LHC, at NNLO QCD accuracy.
Such a final state requires the use of a flavour sensitive jet algorithm to define an IR-safe fixed order prediction beyond NLO QCD.
We computed cross sections and differential distributions using the flavoured modification of the anti-\kt jet algorithm proposed in Ref.~\cite{Czakon:2022wam}.
We compared its output for three different values of the tuneable parameter $a$ against predictions by the flavour-\kt jet algorithm~\cite{Banfi:2006hf}, where we observed a 50\% difference for the integrated cross section, coming from the small \bbbar-distance region.
The scale band is increased accordingly, and can be understood as a consequence of sensitivity to gluon to $b$-quark pair splittings.

We compared our theoretical predictions for the exclusive cross section, where we require exactly two $b$-jets and no other jets in the final state, obtained with the flavoured anti-\kt jet algorithm, against the measurement by the CMS collaboration~\cite{CMS:2016eha}.
We found good agreement for all values of the considered values for the $a$-parameter.
The NNLO QCD corrections are significant, and their inclusion substantially improves the agreement with the data.

We showed differential observables that are particularly sensitive to the choice of the jet algorithm.
The dominant differences between the flavoured \kt and anti-\kt algorithm can be attributed to the region of small angular separation between the $b\bar{b}$-pair.
This region shows also an enhanced sensitivity to the $a$ parameter of the flavoured anti-\kt algorithm.
The determination of an optimal value for $a$ for this process is left for future work.

Finally, we studied the genuine impact of the full NNLO QCD corrections by comparing the NNLO calculation against an improved NLO prediction, that is obtained by merging the $pp\to\wbb$ and $pp\to\wbbj$ NLO QCD calculations, using the exclusive sums technique.
We found that the NLO-merged prediction captures a significant amount of the NNLO QCD corrections, and at the same time shows improvements on the scale dependence, both at the level of integrated cross sections and differential distributions.
Still, the NLO-merged predictions do not capture entirely the NNLO effects and, therefore, the complete NNLO QCD corrections are indispensable.

\acknowledgments
The authors would like to thank Micha\l{} Czakon for making the \textsc{Stripper}
library available to us, and Simon Badger and Alexander Mitov for many useful discussions.
This project received funding from the European Union's Horizon 2020 research and innovation programmes
\textit{New level of theoretical precision for LHC Run 2 and beyond} (grant agreement No 683211),
and \textit{High precision multi-jet dynamics at the LHC} (grant agreement No 772009).
HBH was partially supported by STFC consolidated HEP theory grant ST/T000694/1.
SZ gratefully acknowledges the computing resources provided by the Max Planck Institute for Physics and by the Max Planck Computing \& Data Facility.
AP is also supported by the Cambridge Trust and Trinity College Cambridge.
RP acknowledges the support from the Leverhulme Trust and the Isaac Newton Trust,
as well as the use of the DiRAC Cumulus HPC facility under Grant No.\ PPSP226.

\bibliographystyle{JHEP}
\bibliography{wbb_nnlo}

\end{document}